\documentclass[11pt,twoside]{article}

\usepackage{asp2006}
\usepackage{epsf}
\usepackage{psfig}
\usepackage{lscape}
\usepackage{amssymb,graphicx,amsmath}

\begin{document}
\title{Models for jet power in elliptical galaxies: support for rapidly spinning black holes}   %%% Fill in title
\author{R. S. Nemmen$^1$, R. G. Bower$^2$, A. Babul$^3$, and T. Storchi-Bergmann$^1$ }   %%% Fill in author names
\affil{$^1$Instituto de F\'isica, Universidade Federal do Rio Grande do Sul (Brazil) \\ $^2$Department of Physics, Institute for Computational Cosmology, Durham University (UK) \\ $^3$ Department of Physics \& Astronomy, University of Victoria (Canada)}

\begin{abstract} %%% Abstract to run on from here.
Recently, Allen et al. measured a tight correlation between the Bondi accretion rates and jet powers of the nuclei of nearby X-ray luminous elliptical galaxies. We employ two models of jet powering to understand the above correlation and derive constraints on the spin and accretion rate of the central black holes. The first is the Blandford-Znajek model, in which the spin energy of the hole is extracted as jet power; the second model is an hybrid version of the Blandford-Payne and Blandford-Znajek processes, in which the outflow is generated in the inner parts of the accretion disk. We assume advection-dominated accretion flows (ADAF) and account for general relativistic effects. 
Our modelling implies that for typical values of the disk viscosity parameter $\alpha \sim 0.01 - 1$ the tight correlation implies the narrow range of spins $j \approx 0.7 - 1$ and accretion rates $\dot{M}(R_{\rm ms}) \approx (0.01 - 1) \dot{M}_{\rm Bondi}$. Our results provide support for the ``spin paradigm'' scenario and suggest that the central black holes in the cores of clusters of galaxies must be rapidly rotating in order to drive radio jets powerful enough to quench the cooling flows.
\end{abstract}

%%% MAIN BODY OF TEXT GOES HERE. CONSULT "INSTRUCTIONS FOR AUTHORS USING
%%% LATEX2E MARKUP", SECTIONS 2.3-2.6 FOR HELP WITH EQUATIONS, FIGURES,
%%% AND TABLES.

%\section{}   %%% Top level section head (remove "%" symbol)
%\subsection{}   %%% Second level section head (remove "%" symbol)
%\subsubsection{}   %%% Lowest level section head (remove "%" symbol)
%\section*{}    %%% Unnumbered top level section head (remove "%" symbol)
%\subsection*{}   %%% Unnumbered second level section head (remove "%" symbol)

\section{Introduction}

%The extragalactic jets launched from radio galaxies are some of the most energetic phenomena in the universe, and their nature has been challenging our theoretical understanding of the physical processes involved since they were first observed, more than thirty years ago. It is thought that the radio jets are accelerated and collimated near the central supermassive black hole, by the large-scale magnetic fields anchored in the accretion disk (e.g., \citealt{blandford02}).

Recently, \citet{allen06} (hereafter A06) measured a tight correlation between the Bondi accretion rates $\dot{M}_{\rm Bondi}$ and jet powers $P_{\rm jet}$ of the nuclei of nine nearby, X-ray luminous elliptical galaxies using \textit{Chandra} X-ray observations. 
%This correlation implies that a significant fraction of the energy associated with the material entering the accretion radius is converted to the energy carried by  the relativistic jets. 
It is imperative to understand how this correlation is established and what constraints on the properties of the central black holes can be derived from it. This is the goal of the present work.

\section{Models for the jet power: constraints on the black hole spin and accretion rate}

The origin of jets is likely to be related to the spin of the central black hole (e.g., \citealt{blandford02}). 
With this in mind, we employ two physical models of jet production which relate the spin $j$ ($\equiv a/M_\bullet$) and accretion rate $\dot{M}$ onto the black hole to the observed jet power in order to understand the empirical correlation of A06: the Blandford-Znajek (BZ) model (e.g., \citealt{blandford02}), in which magnetic fields threading the hole extract its rotational energy and drive the jet; and a ``hybrid model'' \citep{meier01} which combines the Blandford-Payne and BZ mechanisms, in which the fields tap energy from the accretion flow and the spinning hole.
We assume that in the nuclei of the observed elliptical galaxies the accretion flow around the black hole is advection-dominated (advection-dominated accretion flows, hereafter ADAF, e.g., \citealt{nemmen06}, see also contribution of F. Yuan in this volume). ADAFs are expected to be associated with the production of radio jets (e.g., \citealt{meier01}). We take into account general relativistic effects not fully appreciated before in such models, in particular the dependence of the radius of the marginally stable orbit ($R_{\rm ms}$) on the black hole spin, and frame-dragging effects that intensify the magnetic field strength. Figure 1 (left) shows the spin dependence of $P_{\rm jet}$ we obtain from the modelling.

Figure 1 (right) illustrates the accretion rates and spins required to reproduce the correlation between $\dot{M}_{\rm Bondi}$ and $P_{\rm jet}$ for $\alpha=0.3$. For typical values of the viscosity parameter $\alpha \sim 0.01 - 1$, we obtain that the tight correlation of A06 implies the narrow range of spins $j \approx 0.7-1$ and accretion rates $\dot{M}_{\rm ms} \approx (0.01 - 1) \dot{M}_{\rm Bondi}$. These results are discussed in more depth in  \citealt{nemmen07}.

\begin{figure*}
\centering
\begin{minipage}[b]{0.51\linewidth}
\includegraphics[width=\linewidth]{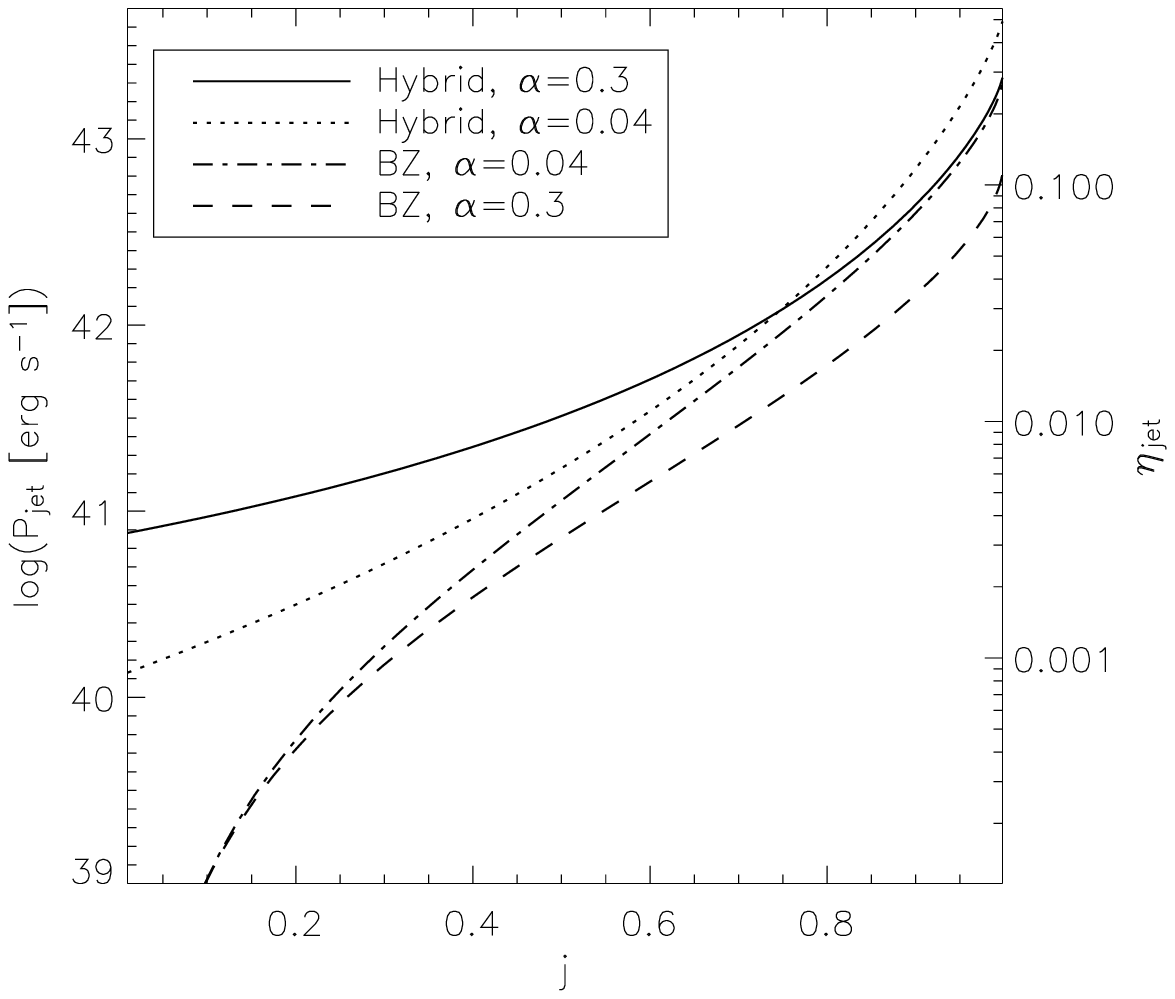}
\end{minipage}
\hfill
\begin{minipage}[b]{0.47\linewidth}
\includegraphics[width=\linewidth]{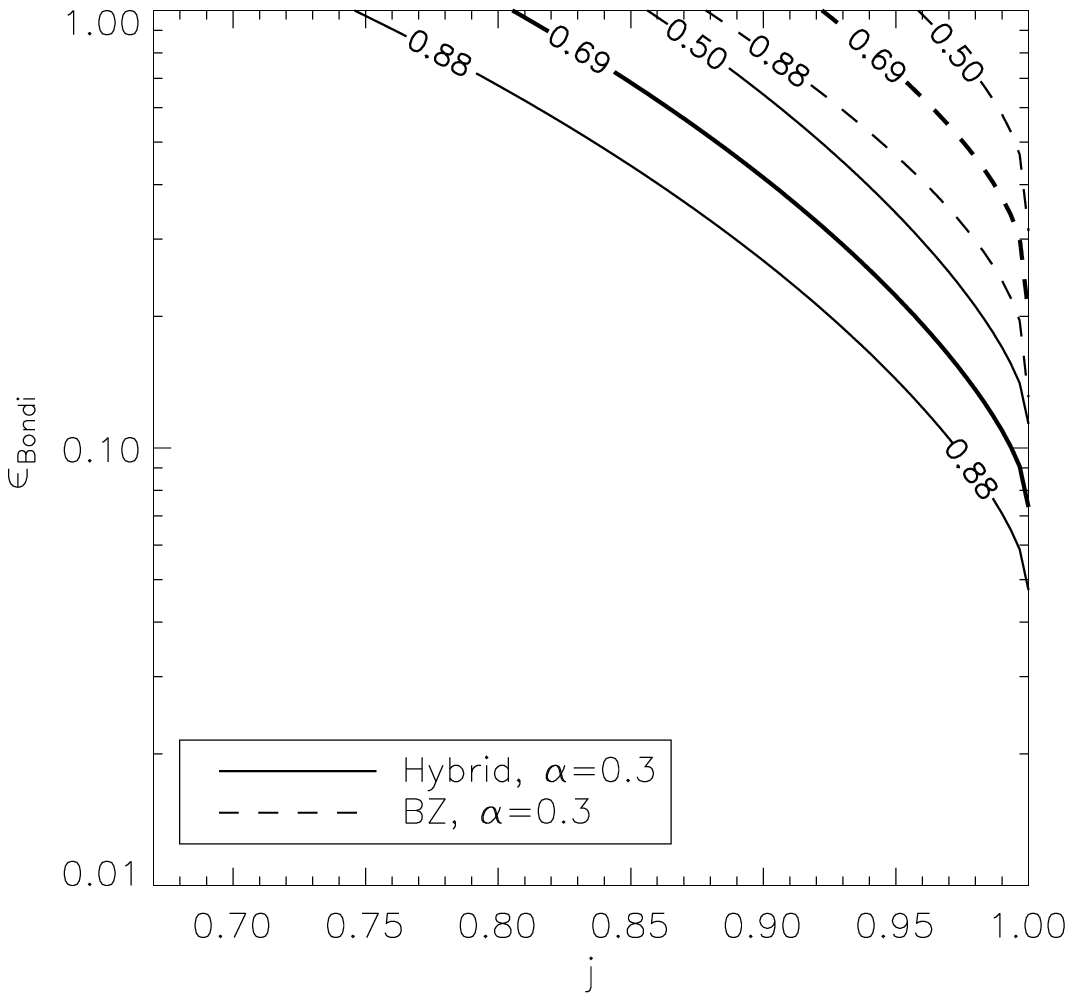}
\end{minipage}
\caption{\textit{Left:} comparison of the spin dependence of $P_{\rm jet}$ predicted by the models for two values of $\alpha$, taking $\dot{M}_{\rm ms}=\dot{M}(R_{\rm ms})=1.6 \times 10^{-3} \; M_\odot \, {\rm yr}^{-1}$. The right axis shows the jet efficiency $\eta_{\rm jet} \equiv P_{\rm jet}/\dot{M}_{\rm ms} c^2$. \textit{Right:} contours of the model parameters $\epsilon_{\rm Bondi}=\dot{M}_{\rm ms}/\dot{M}_{\rm Bondi}$ and $j$ which reproduce the measured values of $A$ ($A=0.69 \pm 0.19$, labels beside each contour) of the correlation of A06, where $\log (P_{\rm Bondi}/10^{43} \, {\rm erg \; s}^{-1}) = A + B \log (P_{\rm jet}/10^{43} \, {\rm erg \; s}^{-1})$ ($B=1$ from the jet models).}
\label{fig:contour}
\end{figure*}

%\acknowledgements %%% Text of acknowledgements runs on after this command.

%%% THE BIBLIOGRAPHY
%%%
%%% CONSULT SECTION 3 OF "INSTRUCTIONS FOR AUTHORS" FOR HOW TO USE NATBIB.
%%% AUTHORS ARE ENCOURAGED TO USE EITHER THE "THEBIBLIOGRAPY" ENVIRONMENT
%%% BY UNCOMMENTING (DELETING THE "%" SYMBOL) THE COMMANDS BELOW, OR BY
%%% USING THE BIBTEX ENVIRONMENT. TO FIND OUT WHICH IS APPLICABLE TO YOUR
%%% CONTRIBUTION, CONSULT THE VOLUME EDITORS FOR YOUR PROCEEDINGS.
%%%

\end{document}